\documentclass[aps,pre,twocolumn,groupedaddress,showpacs]{revtex4}

\usepackage{epsf,graphicx,amssymb,subfigure}

\newcommand{\be}{\begin{equation}}  
\newcommand{\ee}{\end{equation}}  
\newcommand{\ba}{\begin{eqnarray}}  
\newcommand{\ea}{\end{eqnarray}}  
\newcommand{\bi}{\begin{itemize}}  
\newcommand{\ei}{\end{itemize}}  
\newcommand{\bc}{\begin{center}}  
\newcommand{\ec}{\end{center}}  

\begin{document}

\title{Multifractal distribution of spike intervals for two neurons
  with unreliable synapses}

\author{Johannes Kestler and Wolfgang Kinzel}
\affiliation{Institute for Theoretical
Physics, University of W\"urzburg, Am Hubland, 97074 W\"urzburg, Germany}

\date{February 10, 2006}

\begin{abstract}
  Two neurons coupled by unreliable synapses are modeled by
  leaky integrate-and-fire neurons and stochastic on-off synapses. The
  dynamics is mapped to an iterated function system. Numerical
  calculations yield a multifractal distribution of interspike
  intervals.  The Haussdorf, entropy and correlation dimensions are
  calculated as a function of synaptic strength and transmission
  probability.
\end{abstract}

\pacs{05.45.Df, 87.19.La, 05.45.-a, 05.45.Xt}

\maketitle

Neurons communicate via synaptic contacts. When a neuron fires it sends
an electric pulse (spike) along its axon. This spike activates
biochemical processes in the vicinity of a synaptic contact which
change the electrical membrane potential at the neighboring neuron.
However, experiments on synaptic contacts show that this complex
biophysical and biochemical process is not deterministic. 
Any incoming electrical pulse activates the synapse
with some probability, only. In the cortex, transmission 
probabilities between 10\%
and 90\% are reported  \cite{abeles,allen}. 
Although model calculations show that stochastic synapses can
transmit information \cite{fuhrmann} 
it is still an unsolved mystery how a neural
network with unreliable synapses is able to perform reliable
computations.

A quantitative measure of the activity of neurons is the distribution
of interspike intervals. Typically, one observes  broad distributions
which may be described by a simple mathematical approach: Each
neuron is modeled by a stochastic process which is driven by random
uncorrelated synaptic inputs. Hence, usually the effect of unreliable
synapses is modeled  by external uncorrelated noise \cite{tuckwell, gerstner}.

In this paper we investigate the dynamics of two neurons coupled by
unreliable synapses. The synapses are explicitly modeled  by a
Bernoulli process: Any synapse transmits the spike with some
probability $p$ which is independent of the state of the system. 
Our approach allows to calculate the  spike intervals  from an
iterated-function-system (IFS). Our main result is a multifractal
distribution of interspike intervals. The Haussdorf, entropy and 
correlation dimensions are calculated as a function of the synaptic
strength and the probability of synaptic transmission.
We find a  transition between connected and multifractal support 
of the distribution of spike intervals.

In fact, fractal time series of neural spikes which are observed 
in many different biological systems have been related to quantal
neurotransmitter release \cite{lowen}. Our model shows that even a simple on-off
synapse leads to fractal structures of the neural activity.
However,
our simple model makes predictions for the distribution of spike
intervals of two coupled neurons but it does not explain the nature of
fractal time series.

The two neurons are modeled by a leaky integrate-and-fire mechanism 
working above threshold.  In a more general framework, our model is a
system of two identical pulse-coupled oscillators \cite{mirollo}.  
Without synaptic contacts
the neurons are deterministic and oscillate periodically, one obtains
two intervals between the firing times of the two neurons. With
reliable inhibitory synaptic contact, and without any delay of the
synaptic transmission, the two neurons relax into a state of
anti-phase  oscillations with a single spike interval. With unreliable
synapses, however, the system has a broad distribution of spike intervals which
becomes multifractal in some range of the model parameters. 

Each neuron is described by the following differential equation for the
time-dependent membrane potential $V(t)$: 
\be
\label{eins} 
\tau \frac{dV}{dt}=\mu-V(t) 
\ee 
As soon as the potential crosses a threshold value
$\theta$ it is reset to a value $V_{r}<\theta$. In addition it fires,
i.e. it sends a spike to its neighbor which is transmitted with a
probability $p$. If a spike is transmitted it reduces the potential of
the receiving neuron by an amount $J$. For simplicity, we consider
only inhibitory synapses to avoid an introduction of a refractory
time. However, we believe that our main results do not depend on the
details of the model.
  
The neurons are working above threshold, $\theta < \mu$, otherwise
they would not fire at all. Hence the parameter $\mu$ controls the
effect of any mechanism which forces the neurons to fire. Without
synaptic couplings each neuron fires periodically with the period
\be
T= \tau \, \ln\frac{\mu-V_r}{\mu-\theta}
\ee  
Without loss of generality we set $V_r=0$, $\mu=1$ and $\tau=1$, and
in the following we use the parameter $\theta=0.95$ which gives a
period of $T \simeq 2.996 \, \tau$.

\begin{figure}
\includegraphics[width=.35\textwidth]{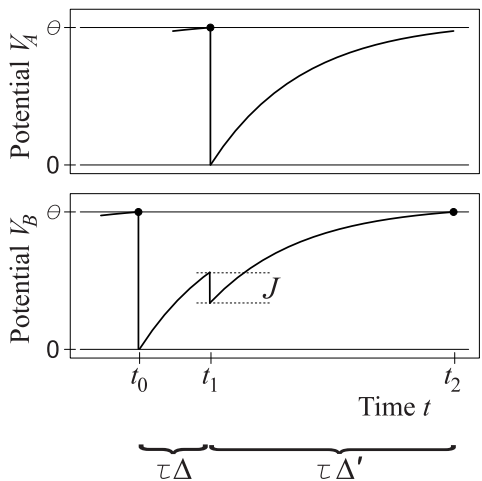}
\caption{Membrane potential of the two neurons. The spike of neuron
  A at time $t_1$ is transmitted to neuron B.}
\label{vt}
\end{figure}

Figure \ref{vt} shows the potential of the two neurons for a typical
situation. At time $t_1$ the neuron A fires and the spike is
transmitted to neuron B resulting in a decrease of the potential by an
amount $J$. The next firing event occurs at time $t_2$. The time interval
between firing events is denoted by $\Delta$. Using the analytic
solution of the differential equation (\ref{eins}) one obtains an
iteration of the spike intervals $\Delta$. For the quantity
$x=\exp(-\Delta)$ the iteration has the form 
\be
x'=f_i(x), \quad i\in\{1,2,3,4,5\}
\ee
where the five functions $f_i$ are selected according to the
transmission probability $p$ and the previous value of $x$. For the situation
of Fig.~1, which occurs with probability $p$ (transmission), one finds
\be
x'=\frac{1-\theta}{x+J}:=f_1(x)
\ee
With probability $1-p$ (no transmission) the sum $\Delta+\Delta'=T$ is
identical to the period of unperturbed oscillations which gives
\be
x'=\frac{1-\theta}{x}:=f_2(x)
\ee
Hence, two simple functions are iterated according to 
probability $p$ of synaptic transmission. The situation becomes
slightly more complicated when neuron A overtakes neuron B, i.e. when one
neuron fires twice before the other one is firing again. This occurs when
the potential $V_B(t_1+)$ becomes negative after neuron A has fired,
that is when $x>1-J$. In this case one has $\Delta'=T$ or
\be
x'=1-\theta:=f_3(x)
\ee
But now $\Delta''$ depends on $\Delta$ and one finds with probability~$p$
\be
x''=\frac{1}{x+J+\frac{J}{1-\theta}}:=f_4(x)
\ee 
and with probability $1-p$
\be
x''=\frac{1}{x+J}:=f_5(x)
\ee
If the synaptic pulse $J$ is larger than $\theta/(2-\theta)$ the same
neuron can even fire more than twice in a row, but we do not consider
such large unphysiological values of $J$.

\begin{figure}

\subfigure[~$J=0.1$]
{  
  \includegraphics[width=.3\textwidth]{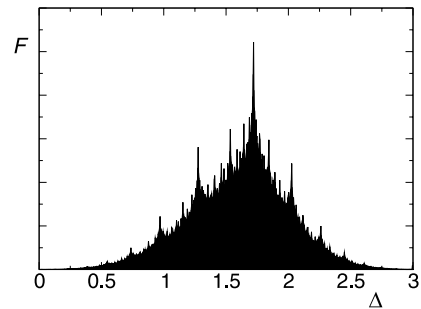}
}
\subfigure[~$J=0.25$]
{  
  \includegraphics[width=.3\textwidth]{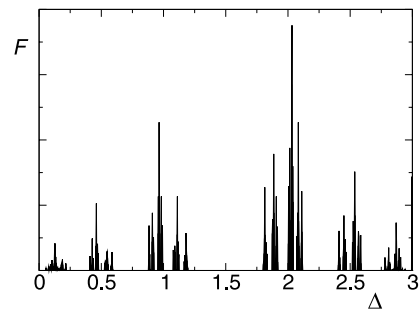}
}

\caption{Histogram of the spike intervals for the transmission
  probability $p=0.5$ and the strength of the synaptic pulse $J=0.1$
  (a) and $J=0.25$ (b). }
\label{f}
\end{figure}

In summary, only five simple functions are iterated to calculate the
distribution of spike intervals $\Delta$. 
It is well known that such a system (IFS) may lead to a fractal
structure of the set of iterated values \cite{barnsley}.
In our numerical simulations
of equations (4) to (8) we have generated about $10^{11}$ spike
intervals for each set of parameters. Figure \ref{f} shows two
histograms of the spike intervals for small and large values of $J$. 
Obviously, the distribution of spike intervals has a complex structure
which we quantify by the R\'enyi dimensions \cite{beck}
\be
D(\beta) = \lim_{\varepsilon \to 0} \frac{1}{\ln \varepsilon}
I(\beta), \quad I(\beta)=
\frac{1}{\beta-1} \ln \sum_{i=1}^r {p_i}^\beta
\ee
Here $\varepsilon$ is the size of the boxes of the histogram and $p_i$
is the normalized number of data points in the box $i$. The sum runs
over all nonempty boxes. For $\beta=1$,  the entropy
$  I(1)=\sum_{i=1}^r p_i \ln p_i$ is calculated.

\begin{figure}
\includegraphics[width=.35\textwidth]{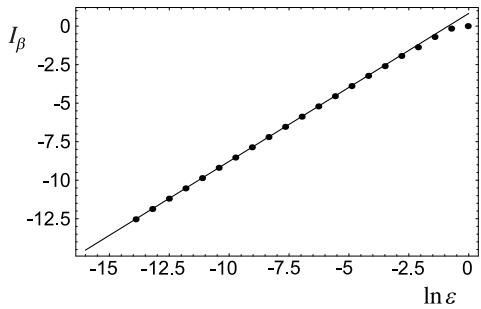}
\caption{The quantity $I_\beta$ as a function of the size
  $\varepsilon$ of the covering boxes (here for $\beta=1$, $p=0.5$ and $J=0.15$). The slope of the figure is an
  estimate of  the R\'enyi dimensions $D(\beta)$ which are shown in
  Fig.~\ref{dim}.}
\label{i-lne}
\end{figure}

We consider three R\'enyi dimensions: The covering or box dimension
$D(0)$ which is usually identical 
to the Haussdorf dimension, the entropy dimension $D(1)$
and  the correlation dimension $D(2)$. Figure \ref{i-lne} shows that a
plot of $I(\beta)$ versus $\ln \varepsilon$ yields a straight line
over several orders of magnitude, hence the corresponding dimension
can reliably be estimated from the slope of this line. In addition, we
checked our results for the correlation dimension by applying the
software package TISEAN to our data \cite{tisean}.

\begin{figure}
\subfigure[]
{
  \includegraphics[width=.35\textwidth]{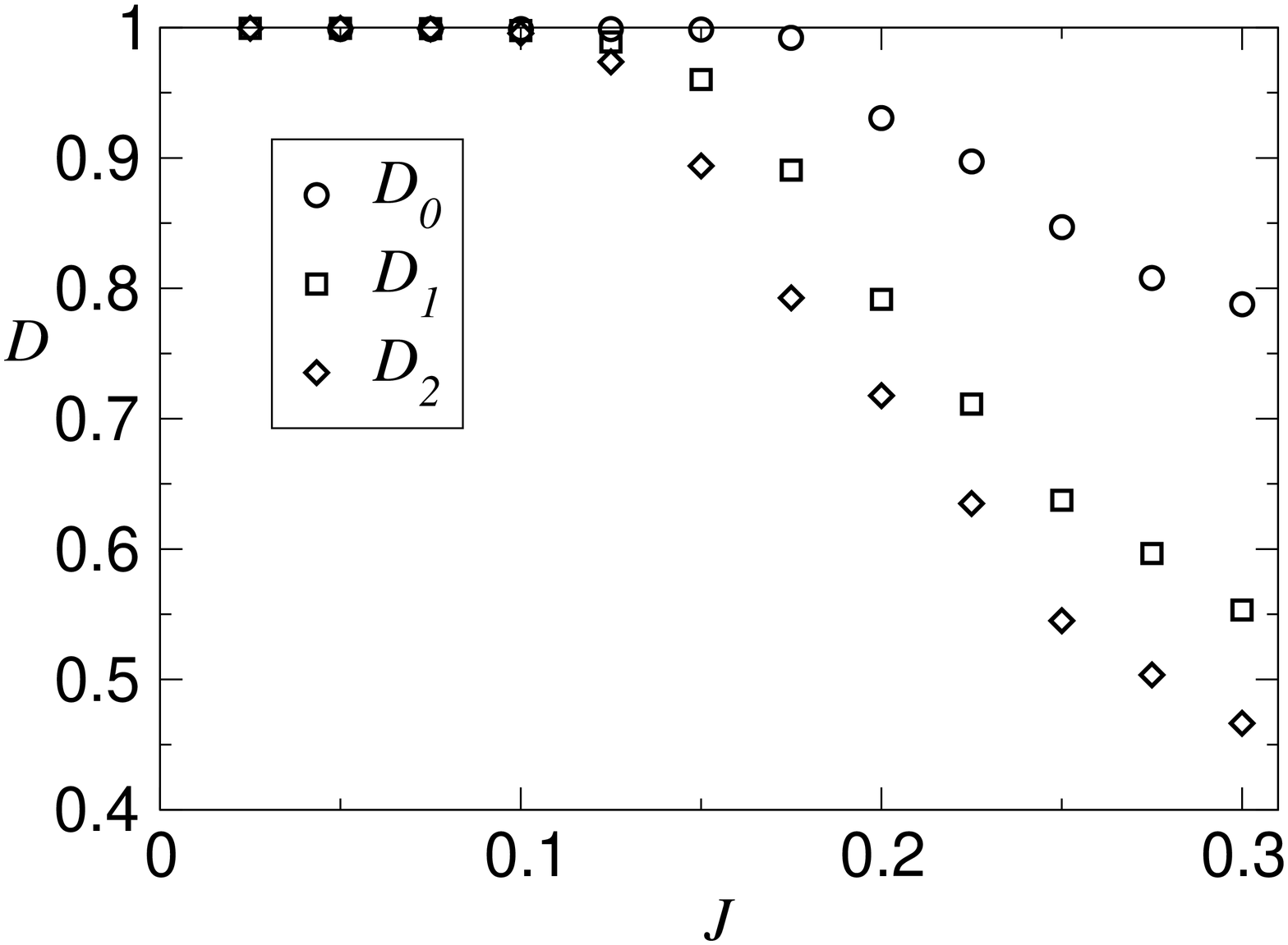}
}
\subfigure[]
{
  \includegraphics[width=.35\textwidth]{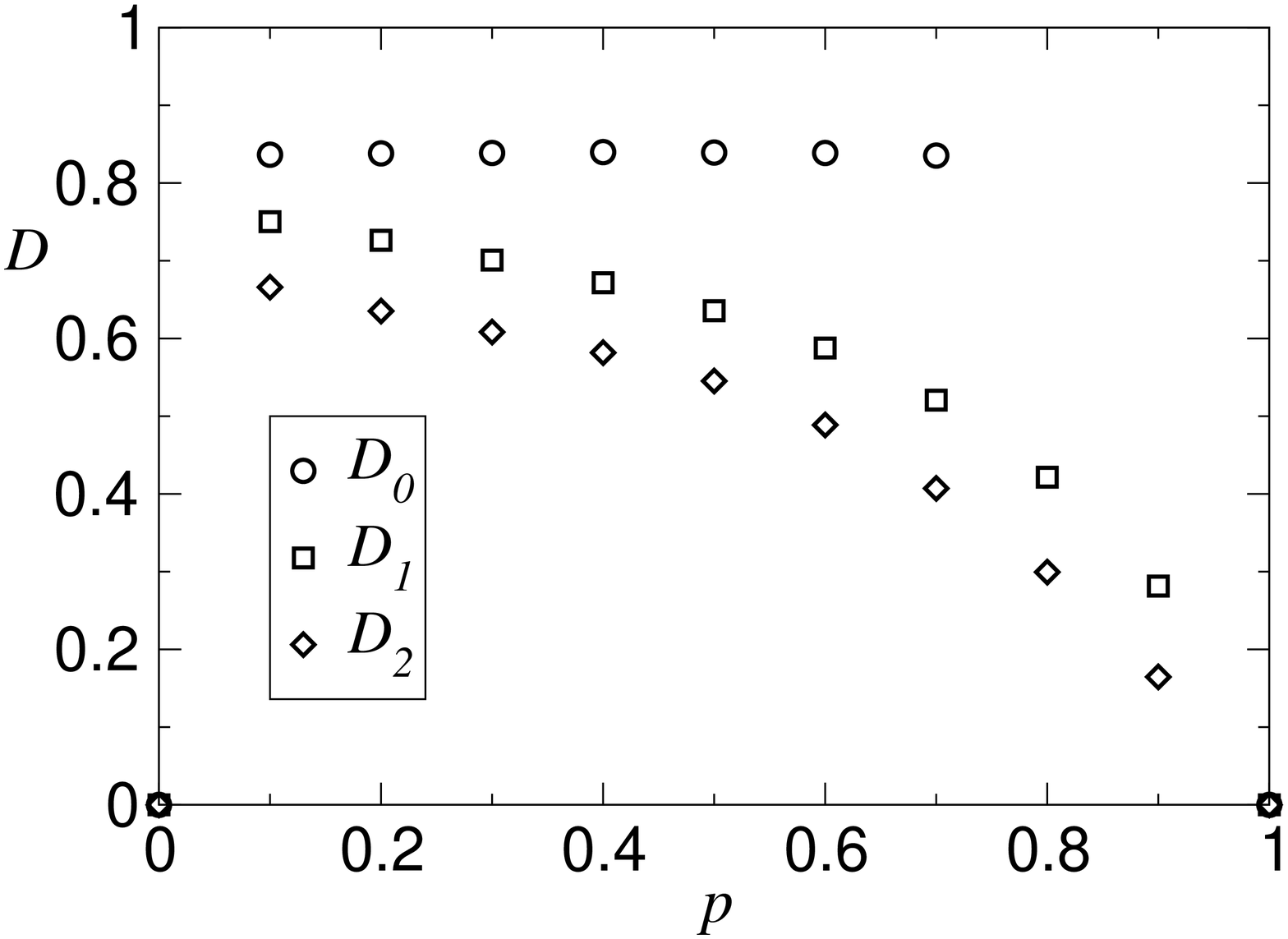}
}

\caption{R\'enyi dimensions (a) as a function of the strength $J$ of the
  synaptic pulse (for $p=0.5$) and (b) as a function of the transmission
  probability $p$ (for $J=0.25$)}
\label{dim}
\end{figure}

The results for the three different R\'enyi dimensions are shown in 
Fig.~\ref{dim}. Of course, our results obey the exact relations $D(2)\le
D(1)\le D(0)$. With increasing coupling strength $J$ and transmission
probability $p$ the three dimensions decrease.  For small values of
$J$ the distribution of spike intervals is smooth, hence one observes
$D(0)\simeq D(1)\simeq D(2)\simeq 1$.  For large values of $J$ the
three dimensions are different, which means that the distribution of
spike intervals is multifractal \cite{beck}. While the covering dimension D(0),
i.e. the structure of the support of the distribution, is insensitive to
the value of $p$, the entropy as well as the correlation dimension
decrease to the value zero in the deterministic limit $p\to 1$. In fact,
for $p=1$,
the distribution of spike intervals is a delta-peak at the fixed point
of $f_1$ which gives $\Delta =
-\ln(-J+\sqrt{4+J^2-4\theta})/2$. Surprisingly, even for $p<1$ the 
distribution has its maximum at this value, a sharp peak, as can be seen from
Fig.~\ref{f}. 

\begin{figure}
\includegraphics[width=.35\textwidth]{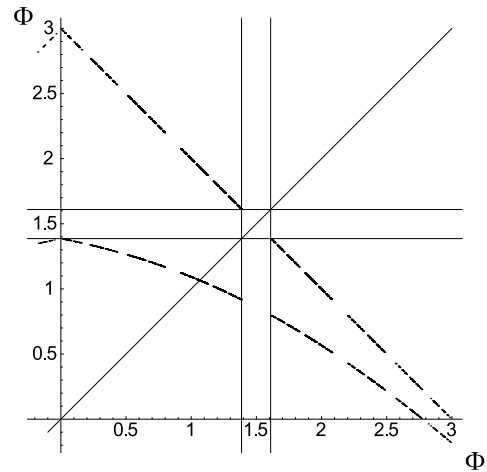}
\caption{The phases $\phi$ of the neurons are iterated by the two
  functions $F_1$ (bottom) and $F_2$ (top) shown by the dashed lines.
The openings of the two functions show the empty intervals in the
distribution of iterated phases.}
\label{phi}
\end{figure}

The results of Fig.~\ref{dim}(a) do not rule out a sharp
transition between a smooth and multifractal distribution of spike
intervals. In fact, for the covering dimension $D(0)$, the transition
point can be found analytically. It is convenient to transform Eq.~(\ref{eins}) 
to $d\phi/dt=1$ where the phase $\phi$ is defined as 

\be
\phi(V)= -\ln(1-V) 
\ee

Now we consider the phase which one neuron occupies after the other
one has fired. After the neuron A has fired it has the phase $\phi=0$,
whereas the other neuron B has a nonzero phase $\phi_i$. If $\phi_i$
is positive it will be neuron B which fires next, namely after the
time $T - \phi_i$.  However, if $\phi_i$ is negative then neuron A
will fire again after the time $T$.  Regardless of which neuron fires,
in both cases we record the phase $\phi_{i+1}$ of the neuron which has
not fired. Given a phase $\phi_i$, the next phase $\phi_{i+1}$ results
by applying one of two mappings depending on whether a spike has been
transmitted at time $t_{i+1}$ or not. These two mappings $\phi_{i+1} =
F_1(\phi_i)$ and $\phi_{i+1} = F_2(\phi_i)$ which describe the
transformation of phases are as follows (see Fig.~\ref{phi}):
\ba
F_1(\phi)&=&-\ln[ \exp(|\phi|-T)+J] ~ \mathrm{(transmission)}\\
F_2(\phi)&=&T-|\phi| ~ \mathrm{(no \; transmission)} 
\ea 
The function $F_2$ just flips the lower interval $[0,T/2]$ to the upper
one $[T/2,T]$. The function $F_1$ maps the complete interval $[0,T]$
to the interval $[-\ln(1+J),-\ln(1-\theta+J)]$. If the maximum of
$F_1$ is smaller than $T/2$, then there exists an interval in the
vicinity of $T/2$ which cannot be reached from outside. In
Fig.~\ref{phi} this interval is indicated by the small square in the
center of the figure. This interval in the center is either flipped by
$F_2$ or mapped to an interval outside of it by $F_1$. This means that
finally any point inside the square will leave it. In addition, no
other point can enter this interval. Hence the distribution of phases
has an opening on this interval. By consecutive iterations of $F_1$
and $F_2$ this opening is distributed on the complete range of phases,
as depicted in Fig.~\ref{phi} by the openings in the functions $F_1$
and $F_2$. This indicates -- but does not prove it -- that the support
of the distribution of spike intervals has a fractal structure,
leading to $D(0)<1$. By these arguments the support 
of the distribution
has a fractal
structure if the maximum of $F_1$ is smaller than $T/2$ which gives a
critical point 
\be J_*=\sqrt{1-\theta}-(1-\theta) \ee 
For $J<J_*$ the distribution fills the complete range of $\phi$
values, while for large values of $J$ the distribution has empty intervals.
Indeed, this value is consistent with the data of Fig.~\ref{dim}(a)
where the covering dimension $D(0)$ deviates from the value $D(0)=1$
at about $J_*=0.1736$. Note, however, that even below $J_*$ the
distribution is multifractal because the values of $D(1)$ and $D(2)$
are still smaller than one. We do not know whether there is a sharp
transition to a smooth structure for small $J$ values or whether the
fractal dimensions $D(1)$ and $D(2)$ 
just come very close to the value one. The data of
Fig.~\ref{dim} do not allow to distinguish between these two
possibilities.

 Our system of two identical pulse-coupled oscillators with random
 on-off synapses is very simplified model of two coupled neurons.  For
 instance, synaptic transmission may be multi-valued \cite{discrete1,discrete2}
 and time-delayed
 \cite{ernst}, and a much better model would include the dynamics of
 ion channels \cite{lowen2}.  However, in any model a random
 uncorrelated process which opens and closes synaptic transmission
 always yields an iterated function system which produces fractal
 distributions of spike intervals depending on the model parameters.
 Up to now, a fractal structure of spike intervals has not yet been
observed. But, to our knowlege, experiments on two interacting neurons under
controlled conditions have not yet been reported, either.  Our model makes 
predictions for such an experiment which may help to clarify the
stochastic nature of synaptic transmission.

\begin{acknowledgments}
We would like to thank Haye Hinrichsen and Georg Reents for useful
discussions.
\end{acknowledgments}

\bibliography{neurons}

\end{document}